\documentstyle[aps,prb,epsf]{revtex}
\newcommand{\Tc}{{\em $T_{c}$}}

\newcommand{\vf}{{\em $v_{f}$}}

\newcommand{\wn}{{\em $cm^{-1}$}}

\newcommand{\wpp}{{\em $\omega_p$}}

\newcommand{\pdc}{{\em $\rho_{d.c.}$}}

\newcommand{\epsi}{{\em $\epsilon_{1} (\omega)$}}

\newcommand{\sig}{{\em $\sigma (\omega)$}}
\newcommand{\sigr}{{\em $\sigma_{1} (\omega)$}}

\newcommand{\mst}{{\em $m^{*} (\omega)$}}

\newcommand{\trw}{{\em $\tau^{*-1} (\omega)$}}
\newcommand{\tuw}{{\em $\tau^{-1} (\omega)$}}
\newcommand{\wsq}{{\em $\omega^2$}}
\newcommand{\wsqrt}{{\em $\omega^{1/2}$}}

\newcommand{\w}{{\em $\omega$}}

\newcommand{\srti}{{\em $SrTiO_{3}$}}
\newcommand{\srru}{{\em $SrRuO_{3}$}}
\newcommand{\casrru}{{\em $Ca_{0.5}Sr_{0.5}RuO_{3}$}}
\newcommand{\srruoooo}{{\em $Sr_{2}RuO_{4}$}}

\begin{document}
\twocolumn[\hsize\textwidth\columnwidth\hsize\csname@twocolumnfalse%
\endcsname
\bibliographystyle{unsrt}
\title{Non-Fermi liquid behavior of \srru\ -- evidence from infrared conductivity.}
\author{P. Kostic, Y. Okada and Z. Schlesinger}
\address{Department of Physics, University of California\\
Santa Cruz, CA 95064}
\author{J. W. Reiner, L. Klein, A. Kapitulnik, T. H. Geballe, and M. R. Beasley.}
\address{Edward L. Ginzton Laboratories, Stanford University\\
Stanford, CA 94305}
\date{March 11, 1998}
\maketitle
\begin{abstract}
The reflectivity of the itinerant ferromagnet \srru\ 
has been measured between 50 and 25,000 \wn\ at temperatures
ranging from 40 to 300 K, and
used to obtain conductivity, scattering rate, and effective mass
as a function of frequency and temperature.  
We find that at low temperatures the conductivity falls unusually slowly as a
function of frequency (proportional to 1/\wsqrt\ ), and at high temperatures it
even appears to increase as a function of frequency in the far-infrared limit.
The data suggest that the charge dynamics of \srru\ are substantially different from
those of Fermi-liquid metals.
\end{abstract}
]

The occurrence of novel phenomena and new physics in correlated-electron systems
is a recurring theme in condensed matter physics.  d-electron
based systems, in particular, present the combined intrigue of a range of 
dramatic phenomena, including superconductivity and itinerant ferromagnetism, and
a tendency towards inscrutability, associated with the fact that key electronic states
are often intermediate between the ideals of localization and itinerancy, which
provide the starting points for most theory.

Ruthenates constitute a class of transition-metal oxides occuring in both
layered\cite{maeno,oguchi} and nearly cubic structures.
Both electronic conduction and magnetic properties are associated with
bands involving Ru-4d orbitals hybridized with O-2p levels,
which manifest
a range of phenomena including superconductivity and magnetism. 
For \srru\ , which exhibits a transition from a paramagnetic to a 
ferromagnetic state at $T_c \approx 150 \, K$,
band-structure calculations are able to reproduce basic aspects
of the magnetic behavior\cite{singh,allen,mazin,santi}.
Specific heat and photoemission 
studies\cite{allen,fujioka}, however, produce results
for density of states and bandwidth
that differ from band theory in manner that indicates the importance
of electron correlation effects.
In addition,
transport studies\cite{allen,lior1,lior2,cao2} of \srru\ show
unusual aspects at low temperature, where a resistivity
minima is observed even while the resistivity remains very low; 
and at high temperature, where the \pdc\
increases nearly linearly with temperature, passing
through the Ioffe-Regel limit without evidence of crossover
or saturation. The latter phenomenon has been regarded as suggestive
of non-Fermi-liquid behavior, and has been used as a phenomenological definition
of ``bad metals'' \cite{emery}.  Evidence that magnetic 
scattering plays a substantial role in the electronic transport
includes the temperature dependence of \pdc\ near \Tc\ ,
where novel critical behavior has been explored by Klein et al.\cite{lior1,lior2}.
\srru\ thus appears to be a correlated system with unusual aspects, 
in which the interplay of magnetism and charge transport, and
the dynamics of strong magnetic scattering, can be fruitfully studied. 
Its relationship to cuprate compounds, which are close to antiferromagnetism,
is also of natural interest.

In this letter we present infrared and optical studies of \srru\  which use 
reflectivity measurements to obtain conductivity, scattering rate and effective 
mass as a function of frequency.  Previously
Bozovic et al.\cite{bozo} and de la Cruz et al.\cite{cruz} have measured
the room temperature reflectivity of ruthenate compounds or alloys,
and Katsufuji et al.\cite{tokura} have studied in detail 
the out-of-plane response of the layered ruthenate, \srruoooo\ .
In the present work we examine for the first time
the temperature dependent low frequency dynamics of \srru\ , 
which is central to understanding the basic nature of charge transport in 
this system.
We find that at low temperature the conductivity tends to follow a power law
of \sigr $\propto$ 1/\wsqrt\ , which is close to the phenomenological fit used for
cuprates\cite{untwinned}, but far from that of conventional metals, where \sigr\ falls
like 1/\wsq\ .  This behavior is expressible in terms of a frequency dependent 
renormalized scattering rate which increases strongly and almost linearly with frequency,
and may be associated with strong interactions
between the current carrying electrons and a magnetic fluctuation spectrum.
Above \Tc\ the scattering rate remains quite high, consistent with
other evidence for the retention of magnetic character at a local scale\cite{dodge},  
and the conductivity appears to increase as a function of \w\ 
at our lowest frequencies. These behaviors represent a substantial departure
from the phenomenology of clean Fermi-liquid metals.  

The samples used in this work are films grown epitaxially on \srti\ substrates.
They are similar to films described previously by Klein et al.\cite{lior1,lior2},
and distinguished by their large residual resistivity ratios and
excellent in-situ RHEED patterns\cite{lior1,lior2}.
For the reflectivity measurements, relatively thick
films are used (e.g., 220, 420 and 1000 nm), and
a Ag reference film is evaporated adjacent to the sample which is mounted in
a circulating Helium cryostat. 
Infrared and optical measurements covering the range from 30 to 25,000 \wn\ are carried out
using both Fourier-transform and grating spectrometers\cite{article}. 
A Kramers-Kronig transform is utilized to obtain 
the real part of the conductivity and dielectric function from the 
measured reflectivity.
Approximately Drude-like extrapolations, which are utilized at low frequency,
do not appreciably effect the data above about 80 \wn\ .
We have experimented with several terminations at high frequency, and
find that while the different approaches can lead to different behavior for
\epsi\ and \sigr\ between 15,000 and 25,000 \wn\ ,
the choice of termination conditions has no effect on the form
of the results below 15,000 \wn\ , 
and does not introduce or eliminate any features in \sigr\ .

\begin{figure}[htbp]
\epsfxsize=3.8in
\centerline{\epsffile{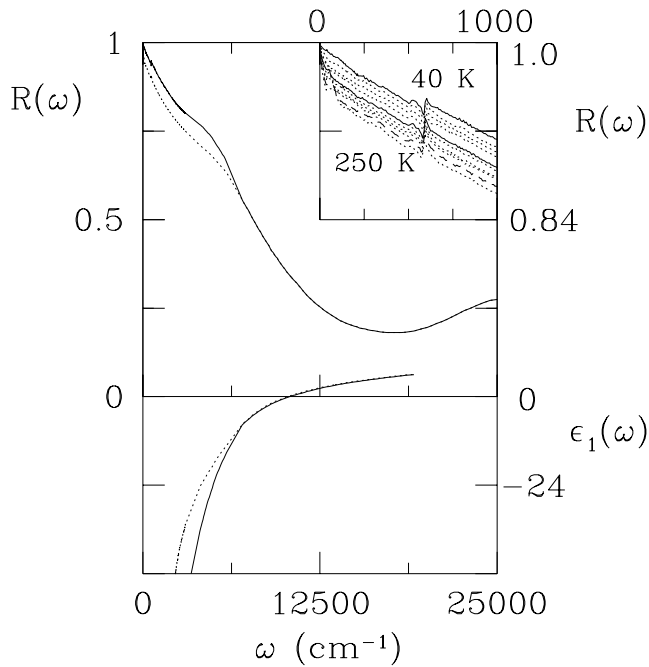}}
\caption{
The reflectivity and the real part of the dielectric function
of a 420 nm \srru\ film are shown at 40 K (solid) and 295 K (dotted).
The inset shows the low frequency reflectivity for the same film at temperatures of
40, 80, 105, 125, 145, 165, 185, 205, 225, 250, and 295 K.
The spectra at 40 K and 145 K are solid, 250 K is dashed and all others
are dotted lines.}
\label{fig1}
\end{figure}

In figure 1 the reflectivity of  \srru\ is shown at 40 and 295 K, 
along with the dielectric function, \epsi\ ,
which crosses zero at about 10,000 \wn . 
The qualitative similarity of the reflectivity of the related ruthenate alloy 
compound \casrru\ to that of a typical
cuprate superconductor has been noted by Bozovic et al.\cite{bozo}.
The temperature dependence of the reflectivity at low frequency is shown 
in the inset to figure 1.  This temperature dependence is closely related to
the temperature dependence of the resistivity, with a higher reflectivity
at low temperature corresponding to a lower resistivity.
 
\begin{figure}[htbp]
\epsfxsize=4.6in
\centerline{\epsffile{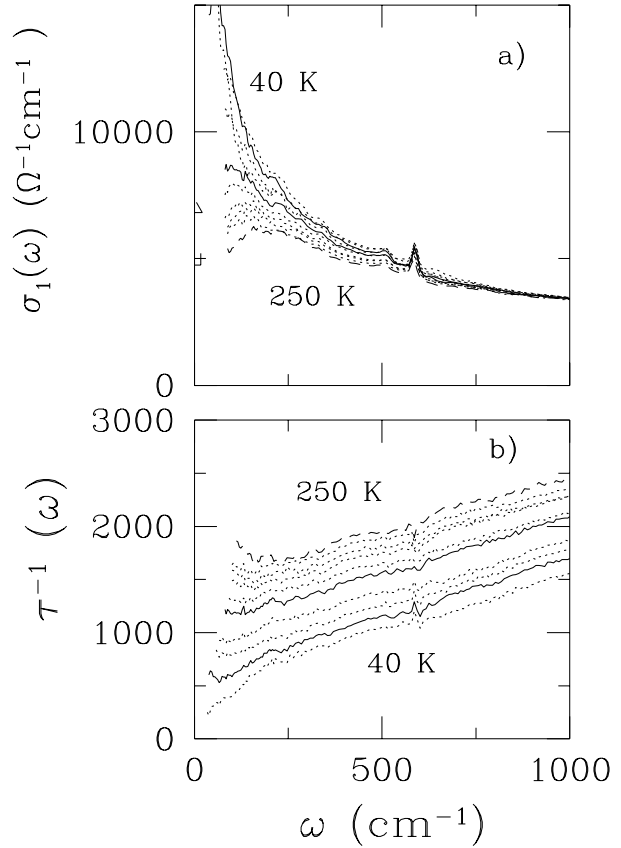}}
\caption{
a) The real part of the conductivity of \srru\ is shown for temperatures of
40, 80 (solid), 105, 125, 145 (solid), 165, 185, 205, 225, and 250 K (dashed).
b) the unrenormalized scattering rate calculated for \wpp\ = 25,000 \wn\ , 
is shown for the same temperature sequence.  The low-T scattering rate 
is well-fit by \wsqrt\ above about $3k_{B} T / \hbar$.
}
\label{fig2}
\end{figure}

From the measured reflectivity we obtain conductivity as a function of
frequency at a dense grid of temperatures from 40 K to room temperature.
In figure 2a we show \sigr\ , along with the unrenormalized scattering rate,
calculated as described below equation (1). 
Our focus in this letter is on the temperature dependent evolution 
of the electron dynamics, as reflected in the overall shape
of \sigr\ .  The narrow peak near 580 \wn\ is associated
direct phonon absorption, which is appropriately subtracted
from \sig\ before performing the scattering rate calculations. 
We note also that over the frequency range of figure 1,
with a putative carrier mass of 4,
the difference in spectral weight (integrated conductivity) 
between the low and high-temperature
\sigr\ spectra corresponds to about 0.2 electrons per ruthenium. 

The data presented here provide a basis with which to test our understanding
of the charge dynamics of \srru\ in particular, and d-level oxide systems in general.  
Several aspects are interesting or unusual.  Although
the low temperature conductivity falls monotonically with frequency,
it does so with a form that is qualitatively different from that of a conventional
metal.  This is emphasized in figure 3a, where the conductivity at 40 K is
shown along with a calculation proportional
to 1/\wsqrt\ , which is much slower than the conventional 1/\wsq\ dependence.
This curve provides a good representation
of the conductivity above about 80 \wn\ .
The conductivity spectra at higher temperatures (145 and 250 K are also shown in
figure 3a) also follow this 1/\wsqrt\ form above a temperature dependent crossover
frequency of order $3k_{B} T$ in each case.
It thus appears that the data can be described in terms of this simple power
law form, augmented by a low frequency cutoff which scales with T 
(allowing \sigr\ to extrapolate to an appropriate, finite d.c. value). 
We know of no theoretical basis on which to understand the proportionality
of \sigr\ to 1/\wsqrt\ .

A second unusual aspect of the data is evident in the high-temperature,
low frequency behavior of \sigr\ .
At 250 K, for example, \sigr\  is relatively flat and
even appears to increase between about 100 and 200 \wn\ . 
This form is reminiscent of the behavior of highly 
disordered systems, 
where the frequency of the maximum of \sigr\ is related to a characteristic
length scale associated with localization.
In the present case, this would imply a length scale of 
roughly 2 or 3 nm at 250 K, however, it is not at all clear what this means 
in this system, which has only a very small amount of disorder
as judged by the very high residual resistance ratio\cite{lior1}.
Perhaps strong inelastic scattering can, in the presence of modest disorder,
lead to such a phenomenology, which could be described as a 
``dynamically induced localization''.
It may be relevant to note here that modest amounts of disorder are associated
with the appearance of a Kondo-like minimum in the resistivity at low 
temperature.\cite{lior1}  Issues related to the position and nature of
the metal insulator phase boundary in various ruthenium-oxide compounds also
remain unresolved and of significant interest.

From the infrared data, and in particular the unrenormalized scattering
rate, shown in figure 2b, one can estimate the electron mean-free-path, $\ell$.
At 145 K, for example, \tuw\ extrapolates to a low frequency value of 
about 1200 \wn\ ($\simeq 2.3 \times 10^{14} \: sec^{-1}$),
which, using calculated values\cite{singh,allen} of \vf\ ,
corresponds to $\ell \simeq$ 0.6 and 1.2 nm for 
majority and minority spin carriers, respectively.
Consistent with other estimates\cite{allen,lior1},
the present values are comparable to the size of a unit cell,
suggesting that this system is close to a 
regime in which ordinary Boltzmann transport theory may not be applicable. 
The low values of the mean-free-path 
may be related to the unusual form of the conductivity as a function of \w\ .

One way in which to view the possible implications of the unusual form of
the conductivity is in terms of a 
frequency dependent scattering rate and mass,
which are calculated from the real and imaginary
parts of \sig\ based on the equation,
\begin{equation}
\sigma (\omega) = {ne^{2} \tau^{\ast} (\omega)\over 
m_b m^{\ast} (\omega) (1+i\omega\tau^{\ast} (\omega))},
\end{equation}
where n is a carrier density, $m_b$ a band mass parameter,
and \trw\ and \mst\ are the renormalized scattering rate 
and mass renormalization parameter, respectively. 
(The unrenormalized scattering rate is 
$\tau^{-1} (\omega) = \tau^{\ast-1} (\omega) / m^{\ast} (\omega)$.)
This approach requires that all the low frequency 
conductivity is associated with a single source of intra-band character,
which should be a good approximation in the present case. 
It provides a natural way to go beyond the simple Drude ansatz,
in which \tuw\ is independent of \w\ and \mst\ $\equiv 1$ .

\begin{figure}[htbp]
\epsfxsize=5.3in
\centerline{\epsffile{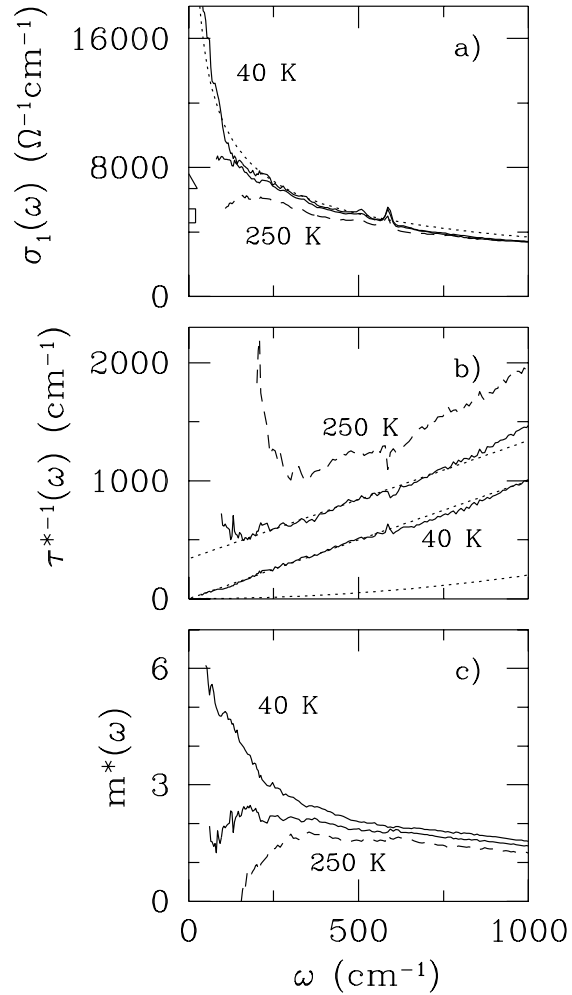}}
\caption{
a) The real part of the conductivity of \srru\ is shown for temperatures of
40 (solid), 145 (solid), and 250 K (dashed).
The dotted curve is a calculation of c/\wsqrt\ , which fits the experimental
conductivity well above a temperature dependent 
crossover frequency of about $3k_{B} T / \hbar$.
The triangle and square on the $\omega = 0$ axis represent estimates
of $\sigma_1 (0)$, at 145 and 250 K respectively, 
from d.c. resistivity measurements.
b) The renormalized scattering rate is shown 
for 40 (solid), 145 and 250 K (dashed). 
The dotted curves are linear fits to \trw\ at 40 and 145 K,
which have the same slope (unity) and intercepts that differ by 340 \wn\ .
The lowest dotted curve is an estimate of the frequency dependence of
\trw\ due to electron-electron scattering in a good metallic ferromagnet.
c) The mass renormalization parameter, \mst\ , is shown 
for 40 (solid), 145 and 250 K (dashed). 
}
\label{fig3}
\end{figure}

The efficacy of this approach for the treatment of both d- and f-electron systems
has been discussed by several authors\cite{allen2,webb,sulew,orenstein1,untwinned,article}.
For a heavy-Fermion system at low temperature, \mst\ is of order unity
at high frequency, and increases to a 
value comparable to the specific heat mass below a characteristic frequency
associated with the development of coherence.
The scattering rate decreases from a high to a low value through
the same frequency range.
One thus observes in the frequency dependence of these quantities 
the crossover from the high scattering rate, incoherent state at high energy,
to the heavy, coherent state at low energy. 
At low temperature and frequency \trw\ is 
proportional to \wsq\ in heavy-fermion systems, and, in general, the leading order
corrections to the scattering rate are 
proportional to $T^2$ and \wsq\ in Fermi-liquid theories.

For \srru\ , the renormalized scattering rate in the range from 50 to
1000 \wn\ , increases linearly with frequency at low T, as shown in figure 3b.
In this regime, the qualitative form of \trw\ is similar to that of an
optimally doped cuprate in the normal state\cite{untwinned}, 
and the magnitude of the scattering
is about 50 \% larger at a comparable temperature.
Further context is provided by comparison to expectations for elemental
ferromagnets.  While the frequency dependent scattering has not been
explicitly determined for such systems, one can infer from d.c. measurements and
theoretical considerations that for good metallic ferromagnets\cite{lior1}
the non-phonon part of the scattering rate is of order $10^{-4} \omega^{2}$ \wn\
(with \w\ in \wn\ )  .  
Shown in figure 3b, this frequency dependence
is dramatically less than that of \srru\ .
The scattering rate of \srru\ is thus very unusual. It differs in its
frequency dependence from that expected for Fermi-liquid metals, and it differs
in its magnitude and frequency dependence from the
inferred scattering rate for Fermi-liquid itinerant ferromagnets.

At higher temperatures, the qualitative form of the conductivity
appears to be even more unusual, and probably beyond the scope of
a frequency dependent scattering description in the sense that \mst\
and \trw\ are neither monotonic nor positive definite at low frequency (fig 3).

In summary, we have used infrared and optical reflectivity measurements to
obtain \sigr\ for \srru\ in both the ferromagnetic and paramagnetic states.
At low temperature \sigr\ falls like 1/\wsqrt\ and \trw\ increases approximately
linearly with \w\ .  At higher temperature, \sigr\  is relatively flat and may
exhibit a maximum in the
far-infrared.  Both regions are difficult to understand within the context of
conventional Eliashberg or Fermi-liquid theory, and suggest the need for
novel theoretical approaches to address the nature of charge response in correlated 
d-level oxide systems such as \srru\ .

Acknowledgments:  The authors acknowledge valuable conversations with
D. L. Cox and J. S. Dodge. 
Work at UCSC supported by the NSF through grant DMR-97-05442.
Work at Standord supported by the DOE through grant 
DE-FG03-94-45528, by the AFOSR through grant F49620-95-1-0039,
and the NSF through the Stanford CMR NSF MRSEC program.

\bibliography{ruo3,zs-short,valence,ir-sc}
\end{document}